\documentstyle[a4,aps,prb,graphics]{revtex}

\begin{document}

\title{Optical phonons in a quarter-filled 1D Hubbard model}

\author{Philippe MAUREL and Marie-Bernadette LEPETIT}

\address{Laboratoire de Physique Quantique, IRSAMC, 118 Route de
  Narbonne, 31062 Toulouse, France} 

\date{\today} 
 
\maketitle

\begin{abstract}

The influence of dispersion-less quantum optical phonons on the phase
 diagram of a quarter-filled Hubbard chain is studied using the
 Density-matrix renormalization group technique. The ground state
 phase diagram is obtained for frequencies corresponding to the
 intra-molecular vibrations in organic conductors. For high
 vibrational modes the system is only slightly affected by the
 electron-phonons coupling. It remains in a L\"uttinger liquid phase
 as long as the electronic repulsion is larger than the polaronic
 binding energy. For low vibrational modes the phase diagram is very
 rich. A noticeable point is the existence of a $4k_F$ CDW phase for
 small values of the correlation strength. For realistic values of the
 electron repulsion and of the electron-phonons coupling constant a
 large phonons-mediated reduction of the L\"uttinger liquid parameter
 $K_\rho$ was found compared to the pure electronic model.

\end{abstract}


\section{Introduction}

Electron-phonons interactions have always attracted a lot of
attention.  Indeed, they are responsible for a large number of phase
transitions, such as BCS supra-conductivity or Peierls transitions.
Among the materials for which electron-phonons coupling are important,
molecular materials constitute a special class.  A trivial statement
is that molecular crystals differ from simple crystals by the fact
that their basic units are molecules. These basic units are therefore
structured systems with a large number of internal degrees of freedom
that may interact with the valence or conduction electrons. In
particular, molecular crystals have two kinds of electron-phonons
interactions. The first one is inter-molecular and the phonons
essentially couple to the electronic structure through a modulation of
the hopping parameter between two molecular sites. This
electron-phonons coupling is responsible for the Peierls transitions
and has been extensively studied. The second type of interaction is
intra-molecular. The totally-symmetric molecular vibrational modes
couple to the electronic structure, essentially through a modification
of the on-site parameters such as molecular orbital energies and
on-site repulsions. In the $60$'s, Little~\cite{Lit} suggested that
intra-molecular vibrations could be responsible for supra-conductivity
in organic conductors, more recently they were proposed as mediators
for supra-conductivity properties in materials such as
fullerides~\cite{c60}.

Indeed in systems such as the organic conductors, there is a broad
spectrum of intra-molecular vibrations ranging from $200 cm^{-1}$ to
$2000 cm^{-1}$ ~\cite{vib1,vib2}, that is between $\sim t/5$ and $\sim
t$, where $t$ is the hopping inter-molecular integral responsible for
the conduction. Moreover it was shown that these vibrations couple
quite well with the electronic degrees of freedom both in the hight
and low vibrational range of the Raman spectrum~\cite{vib4,vib5}. This
fact can easily be understood from simple chemical
considerations. Indeed, the molecules acting as basic units in these
systems share number of characteristics such are being large, strongly
conjugated, and built from pentagonal cycles. These characteristics
allow them to adjust their geometry to their electronic charge at a
low energetic cost, essentially by a modification of the angles in the
pentagonal cycles. This mechanism leads to an important electron
intra-molecular vibration coupling.

Previous studies on correlated electronic systems coupled to
intra-molecular vibrations have essentially explored the two
asymptotic regimes, (i) the weak coupling regime~\cite{voit1,voit2}
that can be treated by perturbative expansion from the electronic
model and (ii) the strong coupling or polaronic
regime~\cite{pol1,pol2,pol3}. It is clear that the study of the
intermediate coupling regime cannot be done by analytic treatments and
requires up to date numerical techniques. The purpose of this work is
to fill this gap and systematically explore the phase diagram of a one
dimensional, correlated, electronic system coupled to intra-molecular
vibrations, both as a function of the coupling constant and the
correlation strength.  The present paper systematically studies the
Hubbard-Holstein model for a one dimensional quarter-filled chain for
two values of the vibration frequency ,$\omega = 0.2 t$ and $\omega =
t$, respectively corresponding to the top and the bottom part of the
organic conductors Raman spectra. For each phase, properties such as
spin and charge gaps, distance dependence of the spin, charge and
singlet correlation functions, etc, are reported.

The next section analyses the Hubbard-Holstein model and develops
the computational choices. The third section reports and discusses the
results and the last section  is devoted to the conclusion.

\section{Model analysis and Computational details}

\subsection{The Hubbard-Holstein model}

The Hubbard-Holstein Hamiltonian associates a Hubbard
Hamiltonian, which includes short range electron correlations, 
with dynamical phonons. The latter are linearly coupled with the electronic degrees of
freedom as in the Holstein model~\cite{holst}.
\begin{eqnarray*}
\label{ham} H &=& H_{e} + H_{ph} + H_{e-ph} \\ {\rm with} &&
\nonumber \\ H_{e}\,\,\,\, & = &
\epsilon\sum_{i,\sigma}n_{i,\sigma} +  t\sum_{i,\sigma}{(c_{i+1,\sigma}^{\dagger}c_{i,\sigma}+
  c_{i,\sigma}^{\dagger}c_{i+1,\sigma})} +
U\sum_{i}{n_{i,\uparrow}n_{i,\downarrow}} \nonumber \\ H_{ph}& =
&\omega\sum_{i}{(b_{i}^{\dagger}b_{i}+1/2)}\nonumber\\ 
H_{e-ph}&=&g\sum_{i}{n_{i}(b_{i}^{\dagger}+b_{i})}\nonumber
\end{eqnarray*} 
where  $c_{i,\sigma}^{\dagger }$, $c_{i,\sigma}$ and $n_{i,\sigma}$ are
the usual creation, annihilation and number operators of electrons of
spin $\sigma$ on site $i$ ($n_i=n_{i,\uparrow}+n_{i,\downarrow}$).
$b_{i}^{\dagger}$ and $b_{i}$ are the intra-molecular phonons creation
and annihilation operators.

From the point of view of the isolated molecule the Hubbard-Holstein
(HH) model tries to mimic the relaxation of the molecular geometry
as a function of the ionicity. Indeed the on-site part of the HH model
can be rewritten as
\begin{eqnarray}
  H_i &=& \varepsilon n_{i,\sigma} + U\, n_{i\uparrow}n_{i\downarrow} + \omega
  \left(b_{i}^{\dagger}b_{i}+1/2\right) +
  g\,n_{i}\left(b_{i}^{\dagger}+b_{i}\right) \\ 
&=& \omega \left[\left( (b_{i}^{\dagger} + n_i{g\over \omega} \right)
    \left(b_{i} + n_i{g\over \omega} \right) + {1\over 2} \right] +
     \left(U - 2{g^2 \over \omega}\right)
    n_{i\uparrow}n_{i\downarrow} + n_i\left(\epsilon - {g^2 \over \omega}
  \right) \label{eq:hi2}
\end{eqnarray}
The above formulation points out the three effects treated in the HH model.
\begin{enumerate}
\item The modification of the molecular orbital energy: $\varepsilon
\longrightarrow \varepsilon - g^2 / \omega$. This effect is very
important (i) in multi-band systems since it strongly affects the relative
filling in the different bands and (ii) in opened systems where electrons
can jump in and out from an external bath. In our case it just changes
the energy reference.
\item The decrease of the effective on-site bi-electronic repulsion:
$U \longrightarrow U - 2g^2/ \omega$. In the strong coupling regime
the effective electron-electron interaction can become attractive due
to the electron-phonons interaction, and two electrons  held
together via molecular vibrations.
\item The displacement of the harmonic oscillator describing the
intra-molecular vibrations as a function of the molecular charge $n_i$:
$\omega \left[b_{i}^{\dagger}b_{i}+1/2\right] \longrightarrow \omega
\left[ \left( b_{i}^{\dagger} + n_i\, g/ \omega \right) \left(b_{i} +
n_i\,g/\omega \right) + 1/2 \right]$. This term mimics the
relaxation of the molecular geometry as a function of the molecule
ionicity. The $\lambda = n_i g^2/\omega$ term acts as an effective
molecular coordinate for which the equilibrium geometry is linearly shifted
from $\lambda=0$, when the molecular site does not carry any 
electron, up to $\lambda = 2 g^2/\omega$, when it carries $2$
electrons. 
\end{enumerate}

One sees from equation~\ref{eq:hi2} that the vibronic molecular states
are coherent phonons states, eigenstates of the shifted harmonic
oscillators. They can be referred as $|n_i, S\!z_i,
\nu_j^{n_i}\rangle$, where $\nu_j^{n}$ is the vibrational quantum
number of the molecule $i$ when it supports $n_i$ electrons,
\begin{eqnarray}
H_i |n_i,Sz_i,\nu_j^{n_i}\rangle &=& 
\left[ n_i\left(\epsilon - {g \over \omega}\right) 
+ \delta(n_i-2)\left(U-2{g^2 \over \omega}\right) + 
\omega\left(\nu_j^{n_i} + {1\over 2}\right)\right]
|n_i,S\!z_i,\nu_j^{n_i}\rangle
\end{eqnarray}
where $\delta$ is the Dirac function.
\subsection{Computational details}

The calculations on the infinite chain are performed using the
infinite system Density Matrix Renormalization Group
method~\cite{dmrg}. The main problem risen by the HH model is the
infinite number of vibronic states on each sites. In order to render
the calculations feasible, the basis set have been truncated to the
lowest vibronic states of each molecular site, that is
$|n_i,S\!z_i,\nu_j\rangle$ such that $\nu_j=0,1$.  This choice is
physically reasonable since (i) we work at $T=0$ and therefore only
the lowest vibronic states are expected to be involved, (ii) the
molecules form well defined entities that are only perturbatively
modified by the presence of their neighbors. The drawback of this choice
is that we directly work in the vibronic basis set and the truncation
of the basis set destroys any further possibility to separate the
electronic degrees of freedom from the vibrational ones.  We will see
later, from a wave function analysis of the different phases, that the
truncation does not affect the results of our calculations as long as
we are not too close to a phase transition. 

In the phase diagram explorations we kept 100 states per renormalized
block, while in the properties calculations 256 states were kept. The
charge and spin gaps where computed using a double extrapolation (i) on
the system size and (ii) on the number of states kept $m$. Typically
the extrapolations over $m$ were done from three DMRG calculations with
respectively $100$, $150$ and $256$ states kept. The maximum number of
sites is $84$ and the correlation functions presented below are done
for this chain length. In some cases, exact diagonalizations on small
clusters have also been performed in order to better analyze either
the wave function or the energy spectrum.

\section{Results}

 The present work explores in a systematic and unbiased way the whole
range of electron-phonons coupling regime (from $g/t=0$ to $g/t=1.25$) as
well as the whole range of correlation strength (from $U/t=0$ to
$U/t=16$), for two values of the phonons frequency: $\omega= 0.2t$ and
$\omega=t$.  These frequencies have been chosen in order to
approximatively correspond to the low and high part of the Raman
spectrum of the $TTF$~\cite{vib3}, $TMTSF$~\cite{vib1} or $M(dmit)_2$~\cite{vib2}
molecules. The adiabatic regime  corresponds to the parameter
domains $\omega/t \ll 1$ and $g/\omega \ll 1$, and the strong coupling
regime  corresponds to $g/\omega \gg 1$ and $\omega/t \gg 1$.

\subsection{$\omega = 0.2t$}

\begin{figure}[h]
%\vspace{1cm}
\centerline{\resizebox{6cm}{6cm}{\includegraphics{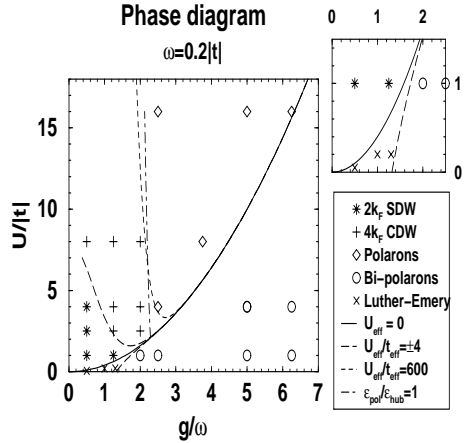}}}
\vspace{5mm}
\caption{Phase diagram of the Hubbard-Holstein model for $\omega =
0.2t$. The solid line corresponds to $U-2g^2/\omega=0$, the
long-dashed line corresponds to $U_{e\!f\!f}/t_{e\!f\!f} =
\left(U-2g^2/\omega\right)/t\exp(-g^2/\omega^2) = \pm 4$, the dashed line
corresponds to $U_{e\!f\!f}/t_{e\!f\!f} = 600$. The dotted-dashed line
correspond to equal phonons polarization  and  electronic
energies (as defined in reference~\cite{pol3}). The inset blows up the small 
$g/\omega$ and small $U/t$ part of the graph.}
\label{fig:diagph02}
\end{figure}

Figure~\ref{fig:diagph02} reports the phase diagram for $\omega=0.2t$
as a function of $g/\omega$ and $U/t$ which are, with $\omega/t$, the three relevant parameters in the HH model (see eq.~\ref{eq:hi2}). 

Five different phases have been characterized: two insulating phases
where the electrons are strongly localized, a polaronic phase
(diamonds) and a bi-polaronic phase (circles) and three metallic
phases (crosses, stars and plus). These phases were characterized
using the following set of tools \\
(i) the variation of the energy with the number of sites, \\
(ii) the spin and charge gaps, \\
(iii) the spin, charge and singlet-singlet correlation functions, \\
(iV) density matrices at the central sites.

\subsubsection{The  bi-polaronic phase}
\begin{figure}[h]
\centerline{\resizebox{6cm}{6cm}{\includegraphics{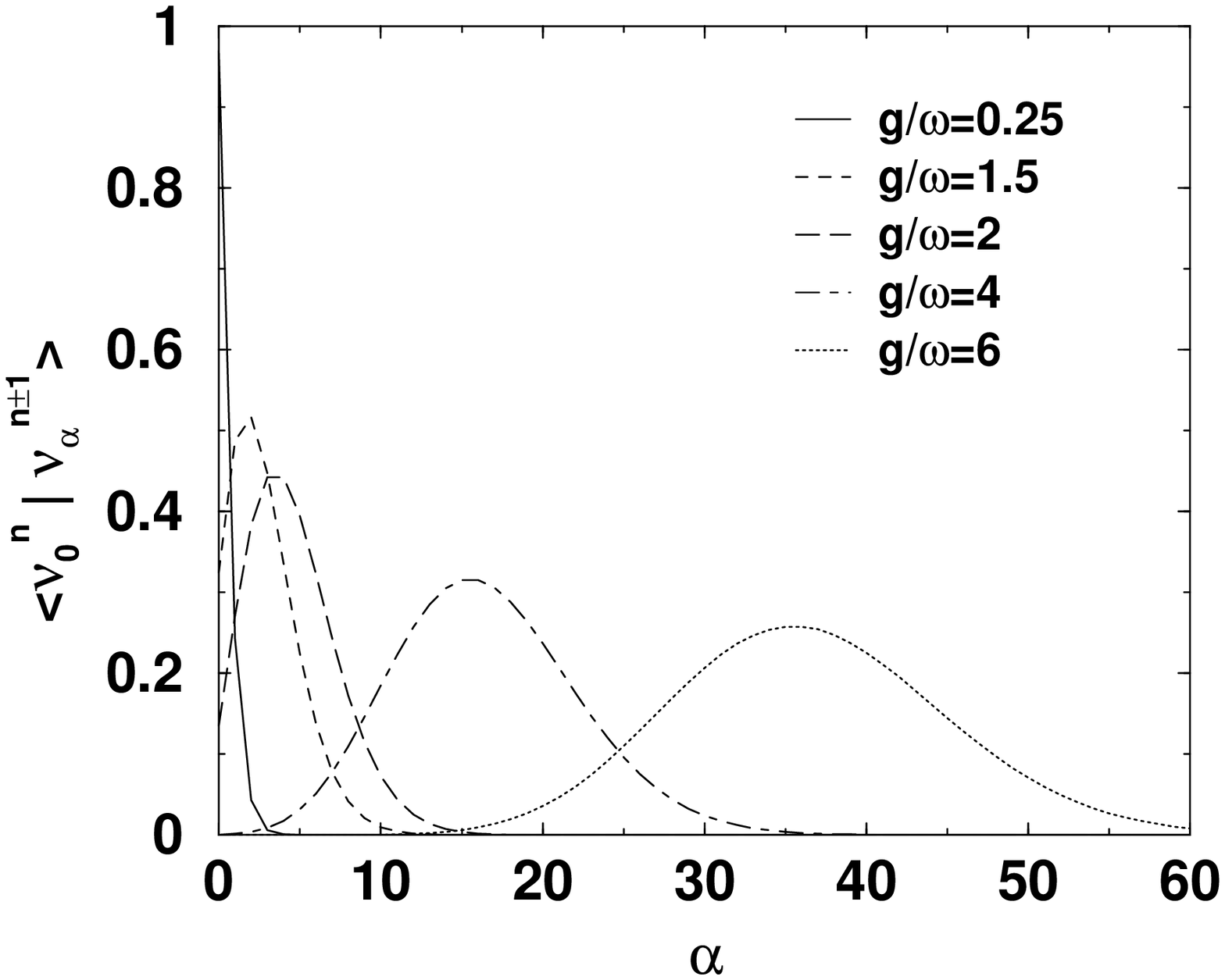}}}
\vspace{5mm}
\caption{Overlap between the ground-state of the undisplaced harmonic
oscillator and the eigenstates of the displaced oscillator for
different values of the displacement $g/\omega$.}
\label{fig:fc}
\end{figure}

In the strong coupling regime, $U_{e\!f\!f} = U -2\omega
\left(g/\omega\right)^2$ becomes negative and a bi-polaronic phase was
found. As expected, the attractive character of $U_{e\!f\!f}$ strongly
couples the electrons in pairs. For instance, for $U/t=4$, $g/t=1$ and
$84$ sites ($U_{e\!f\!f} = -6t$, $g/\omega=5$), the probability of
having a lonely electron on the central site is smaller than
$10^{-12}$. The electron pairing induced by the intra-molecular
vibrations is very strong, however, due to the Franck-Condon factors,
this phase does not correspond to the singlet super-conducting phase
but rather to a localized bi-polaronic phase. Let us suppose that an
electron on a site $i$ would like to hop on a neighboring $j$,
omitting the spin degree of freedom the hopping term can be written as
\begin{eqnarray}
 a^\dagger_{j}a_{i} \;
|n_i,\nu_\alpha^{n_i}\,;\,n_{j},\nu_{\beta}^{n_{j}}\rangle &=& 
t \, |n_i-1,\nu_\alpha^{n_i}\,;\,n_j+1,\nu_\beta^{n_{j}}\rangle \\
&=& t \, \sum_{\gamma,\delta} 
\langle \nu_\alpha^{n_i} | \nu_\gamma^{n_i-1}\rangle \;
\langle \nu_\beta^{n_j} | \nu_\delta^{n_j+1}\rangle  \;
|n_i-1,\nu_\gamma^{n_i-1}\,;\,n_{j}+1,\nu_\delta^{n_{i+1}}\rangle
\end{eqnarray}
 that is the hopping integral between the molecular vibronic ground
states is rescaled by the product of the Franck-Condom factors on the
two sites $i$ and $j$, $\langle \nu_0^{n_i} |
\nu_0^{n_i-1}\rangle \; \langle \nu_0^{n_j} |
\nu_0^{n_j+1}\rangle $. The relaxation energy or self-trapping
energy (due to the vibrations) of the electron pair on a site can be
evaluated as the difference between the vertical ionization
potential (or electron affinity) and the adiabatic one, 
\begin{eqnarray*}
E_{relax}(i) &=& \sum_\alpha \nu_\alpha^{n_i\pm1} \omega \langle
\nu_0^{n_i} | \nu_\alpha^{n_i\pm1} \rangle^2 
\end{eqnarray*}
For $U/t=4$ and $g/t=1$ this relaxation energy is as large as $5.00t$.
Figure~\ref{fig:fc} shows the overlap between the vibrational ground
state corresponding to $n_i$ electrons on site $i$ ($\nu_0^{n_i}$) and
the different vibrational states corresponding to $n_ i\pm 1$ electrons 
($\nu_\alpha^{n_i\pm 1}$). 
One sees immediately that for large or even intermediate
values of $g/\omega$ the overlap between the low energy vibrational
states is very small. The consequence is that the electron hopping
between two neighboring sites is strongly hindered~; either the
transfer takes place towards small quantum-number vibrational states
and the transfer integral is strongly renormalized by the small
Franck-condom factors, 
or the transfer takes place towards large
quantum-number vibrational states and it is hindered by the
vibrational energetic cost. It is clear that the same phenomenon
hinders even more the pair hopping since the displacement is twice as
large and the Franck-Condon factor is squared ($\langle \nu_0^{n_i} |
\nu_0^{n_i\pm2}\rangle = \exp(-2g^2/\omega^2)$).
If one looks at the energy per site as a function of the system size
(or DMRG iteration number) , one sees that it is nearly constant.  The
amplitude of variation between the $4$ sites system and the $84$ sites
system is smaller than $5\times 10^{-12}$ for $U/t=4$ and
$g/\omega=5$, the typical example we have chosen for this phase. This
fact could have been surprising but it is easily understood in the
context of a strongly hindered hopping. Taking into account the
electron pairing and the strong localization one can reasonably
approximate the system wave-function by $\Psi_{BP}$
$$\begin{array}{ccccccccccc}
 &\uparrow\!\downarrow&
          &        &   &\uparrow\!\downarrow&
          &        &   &\uparrow\!\downarrow& \\
\bullet & \bullet &\bullet&\bullet&\bullet&\bullet&
\bullet &\bullet &\bullet&\bullet&\bullet
\end{array}$$
of energy
$$E(\Psi_{BP}) = N_{sites} \;\left({1\over4} U_{e\!f\!f} - {1\over2}
{g^2\over \omega} + {1\over2} \omega \right)$$ The quality of this
approximation can be checked on the energy. The computed DMRG energy
per site differs from $E(\Psi_{BP}) = -3.9$ by at the most $7\times
10^{-12}$ in $t$ units. Exact diagonalization on a $4$-sites system,
where the wave function can be explicitly analyzed, confirms this
result. The projection of the exact ground state on $\Psi_{BP}$ being
larger than $1-10^{-7}$. In fact the ground state of the $4$-sites
system is $4$ times quasi-degenerated and the infinite system ground
state infinitely degenerated. Indeed, in the absence of inter-site
Coulomb repulsion, any choice for the localization of the electron
pairs is equivalent. Even with a $1/r$ Coulomb repulsion one sees that
the infinite system should remain $4$ times quasi-degenerated, the
degeneracy lifting being of the order of magnitude of the rescaled
hopping integral between the low energy vibrational states, that is of
the order of $ t\exp(-g^2/\omega^2)$ ($1.4\times 10^{-11}$ for $U/t=4$
and $g/\omega=5$). The charge gap is therefore exponentially small,
the exact diagonalization of the 4-sites system shows that the first
exited state with a ''real'' gap is a one-boson vibrational excited
state at $\omega$.  The spin gap is of the order of magnitude of
$U_{e\!f\!f}$ ($6.0$ in our example) since it necessitates the
breaking of an electron pair.

\subsubsection{The L\"uttinger liquid phase}
In the weak coupling regime ---~for small values of $g/\omega$~--- up to
the intermediate coupling regime for intermediate values of the
correlation strength, one finds a phase which is essentially a
Tomanaga-L\"uttinger liquid~\cite{Hal,rev}, with parameters slightly rescaled by 
the presence of the vibrations compare to the purely electronic system. 
This result was expected, from continuity from the pure
Hubbard model, and from previous works from Voit and
Schulz~\cite{voit1} (within a renormalization group (Rg) scheme in an 
incommensurate system).  We will further
refer to this phase as the Tomanaga-L\"uttinger (TL) phase.  While
restricted to a very small range of $g/\omega$, for large values of the
correlation strength, it is worth to notice that this phase extends up
to values of $g/\omega$ larger than $1$ for intermediate values of $U$
(see figure~\ref{fig:diagph02}). For very small values of $U$, the
competition between the TL phase and a bi-polaronic phase limits the
extension of the former to a parameter range for which the effective repulsion
remains positive, that is $g/\omega < \sqrt{U/2\omega}$. 
The study of the charge-charge and spin-spin correlation functions
yield that the main effect of the vibrational degrees of freedom is to
rescale the TL liquid parameters ---~in particular the value of
$K_\rho$~--- compared to the purely electronic system. This result is
in total agreement with the conclusions derived by Voit {\it et al.}~\cite{voit1}.

\begin{figure}[h]
\centerline{\resizebox{6cm}{6cm}{\includegraphics{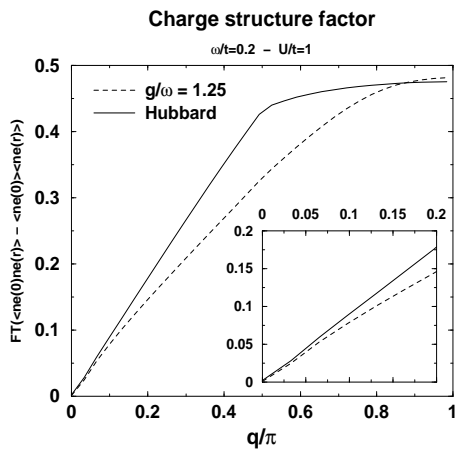}}}
\vspace{5mm}
\caption{Charge structure factor for the pure Hubbard model ($U/t=1$) and
for the HH model ($g/\omega=1.25$, $U/t=1$). The small $q$ part of the
diagram is blown up in the inset.}
\label{fig:tlcorr}
\end{figure}

A simple approach would bring us to think that the value of $K_\rho$
is increased by the vibrational degrees of freedom since the effective
electron-electron repulsion is strongly reduced by the
electron-phonons interactions. Let us see what really comes out of the
calculations.  The correlation functions have been computed on a $84$
sites system with $256$ states kept in the DMRG procedure. They
exhibit the expected behavior as a function of the inter-site distance
with a power law decay, in agreement with the ungaped nature of both
the spin and charge channel.  Figure~\ref{fig:tlcorr} shows the charge
structure factor for the pure Hubbard model as well as the
Hubbard-Holstein one for the $U/t=1$ and $g/\omega=1.25$ set of
parameters.
We can see on figure~\ref{fig:tlcorr} that the numerical calculations
contradict the simple prediction. The structure factor derivative
at $q=0$ is directly proportional to $K_\rho$~\cite{kp}
\begin{eqnarray}
 K_\rho &=& \left.{d\over dq} \left( \pi \int_0^{+\infty} \langle
 \left(ne(0)-\langle ne(0)\rangle\right) \left(ne(r) -\langle ne(r)\rangle\right) \rangle e^{iqr}
 d\!r\right)\right|_{q=0} 
\end{eqnarray}
and is clearly smaller for the $U/t=1$ and $g/\omega=1.25$ set of
parameters than for the $U/t=1$ Hubbard model.  The $K_\rho$ values
computed from the structure factors are $0.79$ for $g/\omega=1.25$ and
$0.89$ for the Hubbard Hamiltonian ---~in total agreement with the
values found in the literature from numerical resolution of the Bethe
ansatz solution~\cite{shultz}.  This unexpected reduction of $K_\rho$
was already noticed by Voit {\it et al.}~\cite{voit1} from Rg 
considerations. One should notice that it cannot be
attributed to the rescaling of the hopping integral since
\begin{equation}
{U_{e\!f\!f} \over t_{e\!f\!f}} ={U-2g^2/\omega \over
t\exp{\left(-g^2/\omega^2\right)}} = 1.31
\end{equation}
 and the corresponding value of
$K_\rho$ is much larger than $0.85$~\cite{shultz}. 

\subsubsection{The $4k_F$ CDW phase}
\begin{figure}[h]
 \begin{minipage}{5cm}
   \centerline{\resizebox{6cm}{6cm}{\includegraphics{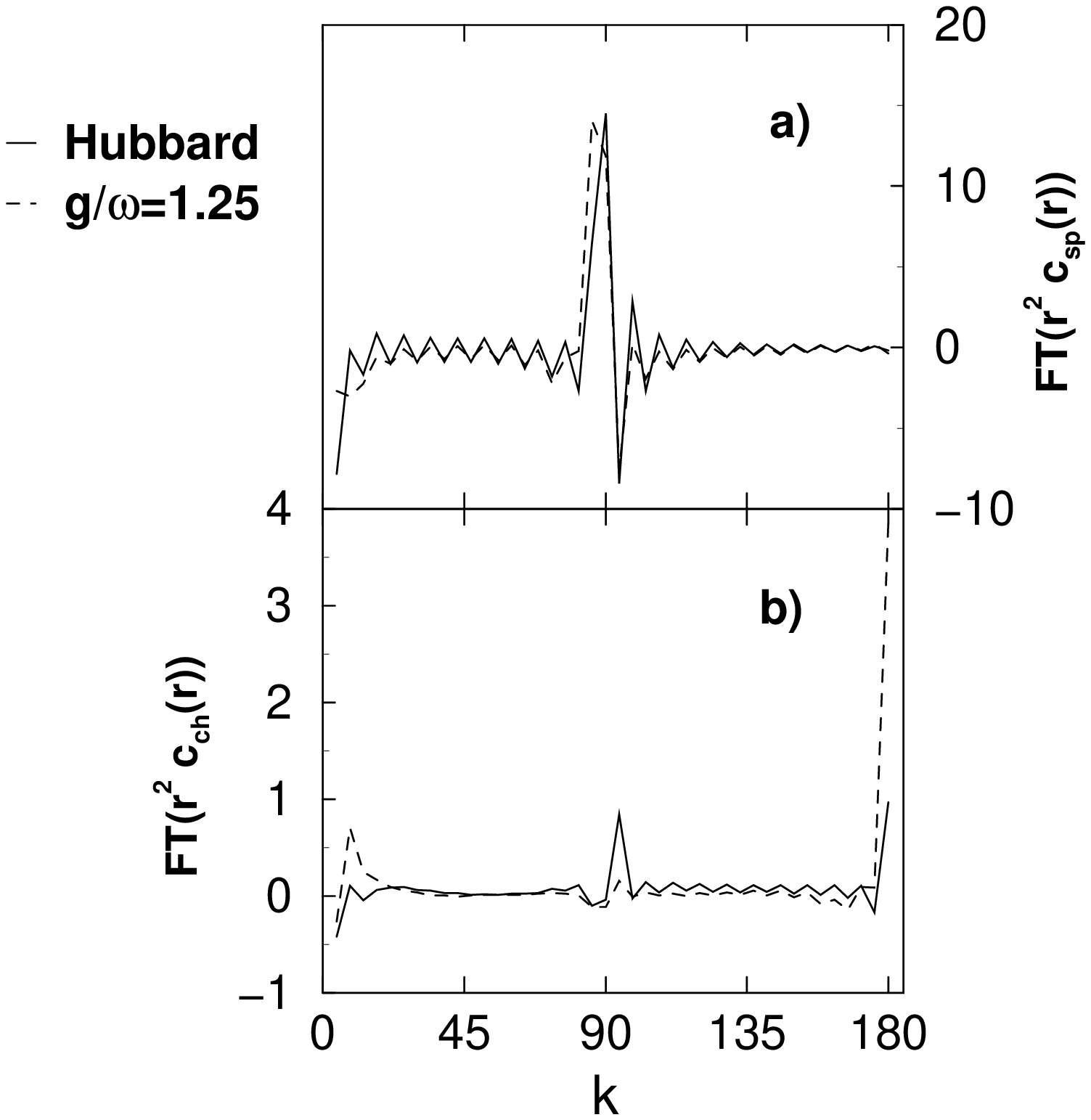}}}
   \caption{Fourier transform of $r^2$ times the charge-charge (b) and
   spin-spin (a) correlation functions for the set of parameters
   $U/t=4$ and $g/\omega=1.25$. The angle is given in degrees.}
   \label{fig:cdwcorr1} \end{minipage} \hspace*{4eM}
   \begin{minipage}{4cm}
   \centerline{\resizebox{5cm}{5cm}{\includegraphics{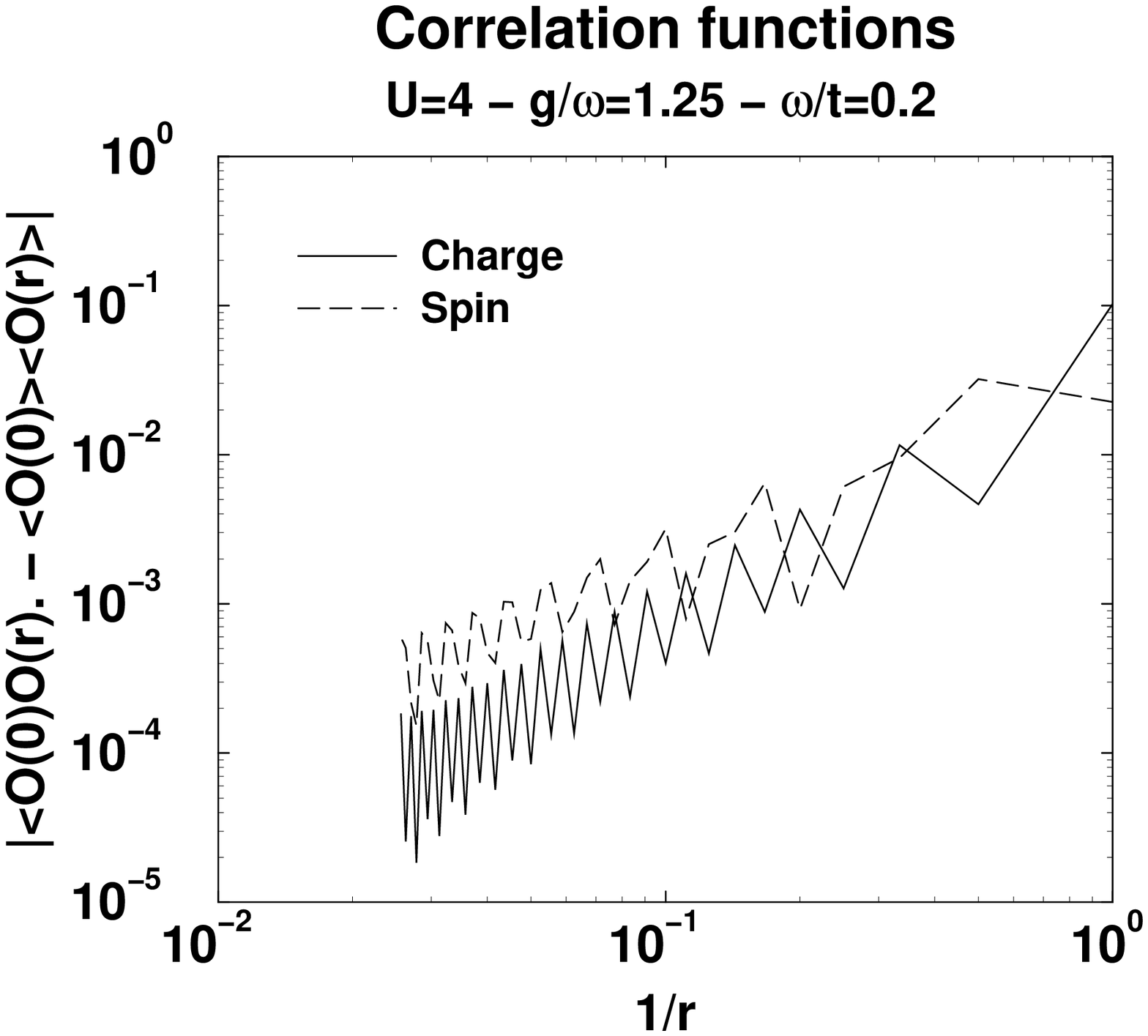}}}
   \caption{Absolute value of charge and spin correlations as a
   function of the inverse of the inter-sites distance ($U/t=4$ and
   $g/\omega=1.25$). }  \label{fig:cdwcorr2} \end{minipage}
\end{figure}

In a regime of intermediate coupling and intermediate to large
correlation strength, we found a phase for which the charge
correlation functions are dominated by $4k_F$ charge density wave
(CDW) fluctuations.  The multiplication of the correlation functions
by $r^2$ ($r$ being the inter-sites distance) allows the elimination,
when the Fourier transform is performed, of the $1/r^2$ term that
contribute at all frequencies. The frequency analysis of
$r^2\times$the correlation function is therefore much clearer, and the
relative importance of the $2k_F$ and $4k_F$ terms enhanced.
Figure~\ref{fig:cdwcorr1} reports this Fourier analysis for $U/t=4$
and $g/\omega=1.25$.  One can see that while the spin-spin correlation
function is change very little compared to the Hubbard model, the
$2k_F$ contribution to the charge-charge correlation function has
disappeared and a strong $4k_F$ contribution has set
place. Figure~\ref{fig:cdwcorr2} reports the same correlation
functions as a function of the inter-sites distance. Both charge and
spin correlation functions decrease as a power law as a function of
the inter-sites distance speaking in favor of a gap-less system in
both the spin and charge channels. The charge and spin gaps have been
computed independently from double extrapolations (i) on the chain
length and on the number of states kept in the DMRG calculations,
i.e. $100$, $150$ and $256$. While the charge channel clearly
extrapolates toward a null gap, in the spin channel the question is
not as clear. Indeed, the gap is found to be $\Delta_\sigma(100)=
8.1\times 10^{-3}$ for a DMRG calculation where $100$ states are kept,
$\Delta_\sigma(150)=8.3\times10^{-3}$ for $150$ states kept and
$\Delta_\sigma(256)=8.6\times 10^{-3}$ for $256$. These gap values are
very small however they do not extrapolate toward a null value when
the quality of the calculation increases. This behavior pleads for a
non null but very small gap in the spin channel. One should note that
this result is not incoherent with the spin-spin correlation function
behavior since such a small gap means a very large coherence length,
of the order of magnitude of $\Delta_\sigma^{-1}$, i.e. the
exponential behavior of the correlation function should take place at
inter-sites distances larger than the chain length.
In the L\"uttinger liquid theory the $4k_F$ phase is supposed to occur
for values of the $K_\rho$ parameter smaller than $1/3$, that is (i)
in very strongly correlated systems and (ii) for values of $K_\rho$
unreachable in the Hubbard model ---~for which $0.5 \leq K_\rho \leq 1$. The
presence of intra-molecular vibrations not only allows the existence
of a $4k_F$ CDW phase, but this phase can be reached for values of the
bi-electronic repulsion as small as $U/t=2.5$ (and most probably even
lower). We computed the $K_\rho$ values from the structure factors and
found that similarly to what happened in the TL phase, the $K_\rho$
value is strongly diminished in comparison to the purely electronic
case. For $U/t=4$ and $g/t=0.25$ the value of $K_\rho$ is $0.56$ while
it is $0.71$ for the $U/t=4$ Hubbard model. One should however note
that the $0.56$ value is  in good agreement with the $K_\rho$
expected for the effective parameters, $U_{e\!f\!f}/t_{e\!f\!f}=16.1$.

\subsubsection{The small polarons phase}
For positive $U_{e\!f\!f}$ but relatively large values of $g/\omega$,
the system is in a small polarons phase (diamonds on
fig.~\ref{fig:diagph02}). Indeed the electrons are no longer paired but are still 
strongly localized. The probability of having two electrons on the central site is
smaller than $10^{-9}$ for $U=8$, $g/t=0.75$ and $82$ sites
($g/\omega=3.75$ and $U_{e\!f\!f}=2.375$). These lonely
electrons remain strongly coupled to the intra-molecular vibrations. 
The on-site vibrational relaxation energy is as large as $2.8t$ and
the rescaled hopping integral between the ground vibrational
states as low as $7.8\times 10^{-7}$.  The ground-state wave-function can be
approximated by the totally localized wave-function $\Psi_P$
$$\begin{array}{ccccccccccc}
 &\uparrow\!& &\downarrow &   &\uparrow& & \downarrow&
      &\uparrow &  \\
\bullet & \bullet &\bullet&\bullet&\bullet&\bullet&
\bullet &\bullet &\bullet&\bullet&\bullet
\end{array}$$ 
and the residual hopping term treated perturbatively. The overlap of
the exact ground-state wave-function of an $4$-sites system with
$\Psi_P$ is as large as $0.99$ for $U/t=8$ and $g/t=0.75$ and even as
large as $0.98$ for $U=4$ and $g/t=0.5$ which is very closed to the
phase transition. Coherently, the DMRG energies per site remains
nearly constant as the system size increases, with a maximal variation 
of $3\times 10^{-7}$ from $ E(\Psi_P)/ N_{sites} = -1/2g^2/\omega+1/2\omega$, 
for $U=8$ and $g/t=0.75$. As for the bi-polaronic phase the
absence of Coulombic inter-site repulsion in our model induces a
strong quasi-degeneracy of the ground-state: $12$ times for the $4$-sites
system but infinitely for the infinite system. Even in the presence of
a $1/r$ repulsion, the system would remain at least twice quasi-degenerated
due to the even sites versus odd sites equivalence with an additional
factor of 4 due to the spin quasi-degeneracy. Capone {\it et
al.}~\cite{pol3} have studied the conditions of existence of
this phase by exact diagonalization on small clusters (4 and 6
sites). They found that the localized polarons phase is stable as long
as the polarization energy defined as $\varepsilon_{pol} = -N_{sites}
\rho (1-\rho) g^2/\omega$ is larger than the electronic energy
$\varepsilon_{elec}$. $\rho$ is the filling of the system. The curve
$\varepsilon_{pol}/\varepsilon_{elec} = 1$ is reported on
fig.~\ref{fig:diagph02} with a dashed-dotted line and fits relatively
well with the phase boundary in the infinite system. In the same
spirit curves of constant $U_{e\!f\!f}/t_{e\!f\!f}$ have been plotted.
This small polarons phase corresponds to very large values of the
$U_{e\!f\!f}/t_{e\!f\!f}$ ratio.

\subsubsection{The Luther-Emery phase}

\begin{figure}[h]
\centerline{\resizebox{6cm}{6cm}{\includegraphics{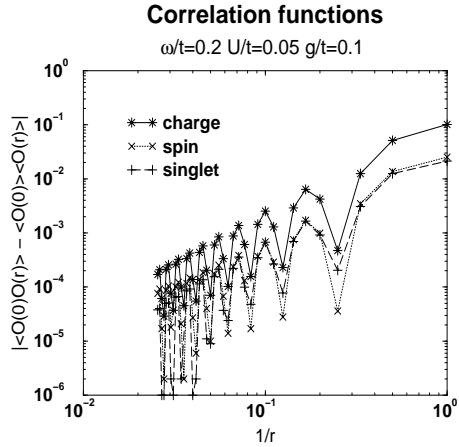}}}
\caption{Absolute value of the charge (stars), spin (crosses) and singlet (plus) correlation 
functions for $U/t=0/05$, $g/t=0.14$ and $\omega/t=0.2$}  
\label{fig:le1corr}
\end{figure}

Finally for negative values of $U_{e\!f\!f}$, but large values of
$t_{e\!f\!f}$ a delocalized phase takes place for which all spin,
charge and on-site singlet correlation functions seem to decrease with
the inter-site distance as a power law. Figure~\ref{fig:le1corr}
reports the absolute values of the correlation functions for
$U/t=0.05$ and $g/t=0.1$, that is $g/\omega=0.5$, $U_{e\!f\!f}=-0.05$
and $t_{e\!f\!f}=0.78$.  In all computed cases ($U/t=0.2$,
$g/\omega=1$ and $1.3$~; $U/t=0.05$, $g/\omega=0.5$) the charge
density fluctuations dominate, however as $U_{e\!f\!f}/t_{e\!f\!f}$
decreases the on-site singlet fluctuations increases compared to the
charge ones.  A direct calculation of the charge and spin gaps yield a
clearly ungaped charge channel and a slightly gaped spin channel.  The
computed spin gaps are $\Delta_\sigma=0.05$ for $U/t=0.05$ and
$g/t=0.1$, and $\Delta_\sigma= 0.06$ for $U/t=0.2$ and
$g/t=0.2$. These very small values are compatible with the the power
law decrease of the spin correlation functions since with such small
gaps, the correlation length is very large ---~of the order of
magnitude of $\Delta_\sigma^{-1}$~--- and therefore the exponential
decay of the correlation function cannot take place at distances
smaller than the order of magnitude of the chain length. One can
therefore identify this phase with a Luther-Emery model~\cite{LE}.
The values of $K_\rho$ extracted from the structure factors are again
smaller than the values of the purely electronic model, with
$K_\rho=0.99$ for $U/t=0.05$ and $g/t=0.1$, and $K_\rho=0.89$ for
$U/t=0.2$ and $g/t=0.2$. Of course, they are as well smaller than the
$K_\rho$ values corresponding to the negative $U_{e\!f\!f}$.

\subsection{$\omega = t$}
\begin{figure}[h]
\centerline{\resizebox{6cm}{6cm}{\includegraphics{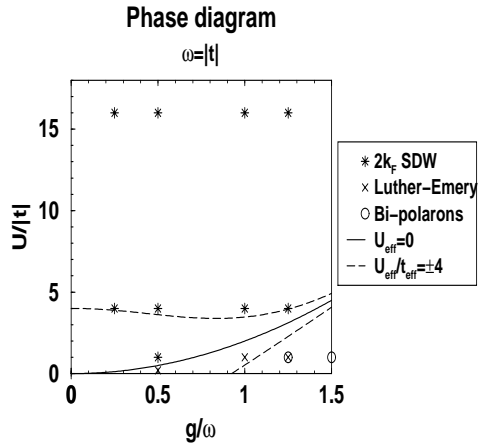}}}
\vspace{5mm}
\caption{Phase diagram of the Hubbard-Holstein model for $\omega = t$. The solid line corresponds to $U-2g^2/\omega=0$.}
\label{fig:diagph1}
\end{figure}

Figure~\ref{fig:diagph1} reports the phase diagram for $\omega=t$
as a function of $g/\omega$ and $U/t$. 

For this intermediate value of the phonons frequency, one finds only
three phases, a slightly perturbed L\"uttinger liquid phase for
$U_{e\!f\!f}>0$, a Luther-Emery phase for $U_{e\!f\!f}<0$ and small
$g/\omega$, a localized bi-polarons phase for $U_{e\!f\!f}<0$ and
larger $g/\omega$.

\begin{figure}[h]
\centerline{\resizebox{6cm}{6cm}{\includegraphics{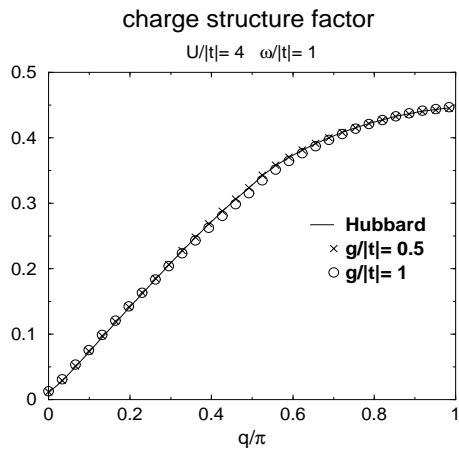}}}
\vspace{5mm}
\caption{Structure factors for $U/t=4$. The solid line corresponds to the pure Hubbard model, the crosses and the circles to the Hubbard-Holstein model with respectively $g/t=.0.5$ and $1$.}
\label{fig:cstu4o1}
\end{figure}

For $U_{e\!f\!f}>0$, the phase is delocalized, dominated by $2k_F$ SDW
fluctuations. Both charge and spin correlation functions decrease as
power laws as a function of the inter-sites distance and both spin and
charge channels are ungaped to numerical accuracy. The structure
factors for the $U/t=4$ Hubbard model and for the Hubbard-Holstein
model with $U/t=4$ $g/\omega=0.5$ and $U/t=4$ $g/\omega=1$ are
reported in figure~\ref{fig:cstu4o1}. As can be seen the HH model is in this
case indistinguishable from the pure Hubbard model and, as a
consequence, the $K_\rho$ values as not sensitive (to numerical
accuracy) to the electron-phonons coupling.  In fact everything goes
as if we were in the adiabatic regime while the model parameters can
be as far from it as $\omega/t=1$ and $g/\omega=1.25$. This is typically a
case where the Born-Oppenheimer approximation is fully valid and the
electronic and vibrational degrees of freedom are nearly independent.

For $U_{e\!f\!f}<0$ one has a  Luther-Emery phase for
small values of $g/\omega$. This phase is very similar to the one
found for $\omega/t=0.2$, with an effective attractive on-site
interaction and large effective hopping integrals. The CDW and the
on-site singlet fluctuations dominate, and become of the same order of
magnitude for $g/\omega=0.5$ and $U/t=0.2$. For larger values of
the electron-phonons coupling, the CDW are larger than the singlet
fluctuations ($g/\omega=1$ and $U/t=1$).  As for the $\omega=0.2$
case, all three correlation functions ---~spin-spin, charge-charge and
singlet-singlet~--- present a power law behavior as a function of the
inter-sites distance for the $84$ sites system. The direct calculation
of the charge and spin gaps yield and ungaped charge channel and a
slightly gaped spin channel with $\Delta_\sigma=0.04t$ for $U/t=1$
and $g/t=1$, as well as  for $U/t=0.2$ and
$g/\omega=0.5$. Again, the smallness of the spin gap explains the
power law behavior of the spin-spin correlation functions at the
computed distances. This time however the values of $K_\rho$ are
larger than in the Hubbard model, in agreement with the attractive
effective on-site interaction. $K_\rho\simeq 1.1$ for both $U/t=1$ 
$g/t=1$ and $U/t=0.2$ $g/t=0.5$. 

\begin{figure}[h]
\centerline{\resizebox{6cm}{6cm}{\includegraphics{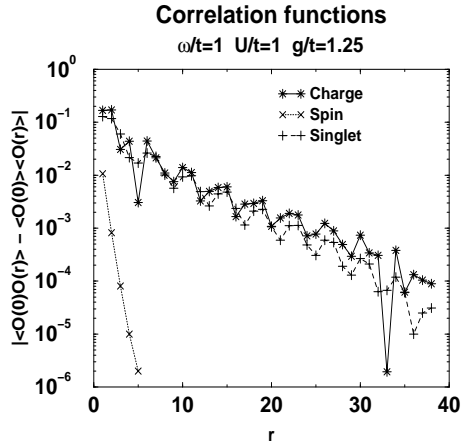}}}
\vspace{5mm}
\caption{Absolute values of the charge (stars), spin (cross) and singlet (plus) correlation functions 
in a semi-log scale. $U/t=1$,
$g/t=1.25$ and $\omega=t$.}
\label{fig:U1g125w1}
\end{figure}

When $g/\omega$ increases ---~but still for negative $U_{e\!f\!f}$~---
one goes from the Luther-Emery phase toward a localized bi-polaronic
phase through a cross-over. Indeed, for $U/t=1$ and $g/\omega=1.5$ the
system is clearly localized in a bi-polaronic phase with the usual
characteristics~: nearly constant energy per site, large projection of
the wave function on $\Psi_{BP}$ ($0.94$ for the 4 sites system),
etc...  For $U/t=1$ and $g/\omega=1.25$ however, the system is in an
intermediate regime. The delocalization is still reasonably large with
an effective hopping of $0.21t$ and a projection of the 4 sites system
wave function on $\Psi_{BP}$ of only $0.72$. In this cross over
regime, the system is gaped both in the spin and charge channels,
with a spin gap of $\Delta_\sigma=0.43$ and a charge gap of
$\Delta_\rho=0.84$.  Figure~\ref{fig:U1g125w1} reports the charge,
spin, and singlet fluctuations. As expected all of them decrease
exponentially with the inter-site distance.  When the electron-phonons
coupling increases, the system localization increases and the wave
function tends toward the totally localized wave function$\Psi_{BP}$.
Coherently, the spin gap increases and is expected to follow the same
variations as $|U_{eff}|=2g^2/\omega -U$, since one needs $|U_{e\!f\!f}|$ 
in order to break the singlet pairing on the sites. In
totally localized systems ($g=+\infty$), the charge channel is
ungaped since in the absence of any delocalization integral, the
different possible choices for the localization of the pairs have the
same energy. The charge gap can therefore be expected to be scaling as
the rescaled hopping integral $t_{e\!f\!f}=t\exp(-g^2/\omega^2)$. That is 
exponentially decreasing when the electron-phonons coupling increases.

\section{Conclusion}
We have investigated the phase diagram of the Hubbard-Holstein
Hamiltonian in a quarter-filled chain for two values of the phonons
frequency. These frequencies have be chosen in order to correspond to
the low ($\omega=0.2t$) and hight ($\omega=t$) frequencies of the
intra-molecular totally-symmetric vibrations of $TTF$, $TMTSF$ and
related molecules from which the organic conductors are built. 

For the $\omega=t$ frequency, the phase diagram is very simple. For a
positive effective on-site interactions
($U_{e\!f\!f}=U-2g^2/\omega>0$), the system is in a L\"uttinger-Liquid
phase with slightly rescaled parameters compared to the purely
electronic system.  For negative effective on-site interactions
($U_{e\!f\!f}<0$) and small electron-phonons coupling, the system is in
a Luther-Emery phase, however the spin gap remains very small and the
coherence length large. The $K_\rho$ parameter is again rescaled
compared to the purely electronic system, in agreement with the
effective model $U_{e\!f\!f}=U-2g^2/\omega$,
$t_{e\!f\!f}=t\exp(-g^2/\omega^2)$. When the electron-phonons coupling
increases, the system goes toward a localized bi-polaronic phase
through a soft cross-over. 
  
At low frequency ($\omega=0.2t$) however, a rich phase diagram have
been found with 5 different phases, some delocalized and some strongly
localized.  For positive effective interactions one has three
different phases. For small electron-phonons coupling and intermediate
coupling for small correlation strength, the system is in a L\"uttinger
Liquid phase. For intermediate electron-phonons coupling, the system is
in a phase for which the $2k_F$ charge fluctuations disappear and the
charge correlation functions have only $4k_F$ CDW fluctuations. For
large electron-phonons coupling, the system is strongly localized in a
polaronic phase. When the effective interaction become negative, one
finds a Luther-Emery phase for small electron-phonons coupling, and a
strongly localized bi-polaronic phase for intermediate to large
couplings. The Luther-Emery phase has a very small spin gap (of the
order of $0.05t$ and the spin-spin correlation functions exhibit a
power law decrease up to at least 60 to 80 sites). Unlike for
$\omega/t=1$, for $\omega/t=0.2$ the delocalized phases do not behave
as would the $U_{e\!f\!f}=U-2g^2/\omega$,
$t_{e\!f\!f}=t\exp(-g^2/\omega^2)$ effective Hubbard model. Indeed the
$K_\rho$ parameter is strongly decreased due to the electron-phonons
interactions while the decrease of the effective repulsion (compared to
the pure Hubbard model) would have led us to expect an opposite behavior.

This surprising decrease of the $K_\rho$ parameters should be put in
context with values of the $K_\rho$ parameters found in photo-emission 
experiments on $TMTSF$ and $TMTTF$ quasi-one-dimensional organic
compounds~\cite{arpes1,arpes2}. Indeed Zwick {\it et al.}~\cite{arpes1} 
found a density of states exponent $\alpha=1/4\left(K_\rho+1/K_\rho -2 \right)$ 
slightly larger than 1 ($K_\rho \leq 0.25)$ in the metallic phase. 
According to the L\"uttinger Liquid theory the
system should be insulating for such large values of $\alpha$.
Tentative explanations have been made by Schwartz {\it et al}~\cite{opt}, 
involving an effective doping of a Mott insulator due to the inter-chain hopping.
On another hand, the electron intra-molecular vibrations interactions
are never considered in these systems. For the low frequency modes of
these systems, the typical energy scale of the electron-phonons
interaction~: $2g^2/\omega$ is of the order of magnitude of one-half
to one-fourth the intra-chain hopping amplitude~\cite{vib4,vib5}, that is in the
intermediate regime of our phase diagram. For such values of the
electron-phonons coupling constants and values of the electronic
repulsion between one to four $t$, the system is in a L\"uttinger liquid 
phase for which the intra-molecular vibrations strongly
decrease the $K_\rho$ parameter compared to its electronic value. It
is clear that the Hubbard model lacks at least nearest-neighbor
bi-electronic repulsion in order to well represent these systems.
It can however be expected that the trend toward a strong reduction of
the $K_\rho$ parameters, observed in the Hubbard-Holstein model, would be similar
with longer range interactions, and that the coupling of the
electronic degrees of freedom with the intra-molecular vibrations
would lead to values of the $\alpha$ parameter unreachable in a
purely electronic model. Considering these trends, it is reasonable to
think the the electron intra-molecular vibrations coupling will have to be 
considered in order to obtain a realistic description of the observed
photo-emission experimental data.

\end{document}